\documentclass[10pt,draft]{dis03}
\usepackage{epsf,amsmath}

\usepackage{graphics}
\usepackage{floatflt}
\usepackage{color}
\usepackage{colortbl}
\newcommand{\Pt}{{P_t}}

\newcommand{\zpj}{``$Z^0+jet$''~}
\newcommand{\gpj}{``$\gamma+jet$''~}
\newcommand{\gzpj}{``$\gamma/Z^0+jet$''~}
\newcommand{\rrr}{\to} 
\newcommand{\pth}{\hat{p}_{\perp}^{\;min}}
\newcommand{\ptgj}{$~\Pt^{\gamma}$ and $\Pt^{jet}~$}

\newcommand{\Ptg}{\Pt^{\gamma}}

\newcommand{\lt}{\!\!<\!\!}

\newcommand{\com}{qg\to q+\gamma}
\newcommand{\ann}{q\bar{q} \to \gamma + g}
\newcommand{\hmm}{\hspace*{-1.3mm}}      
\newcommand{\coltab}{0.97}
\newcommand{\tg}{\tilde{\gamma}}

\suppressfloats[!]

\begin{document}

\vskip-10mm
\title{ON THE POSSIBILITIES OF MEASURING THE GLUON DISTRIBUTION USING
"$\gamma/Z^0$+JET" EVENTS AT TEVATRON RUN~II AND LHC.}

\author{D.V.~Bandurin$^1$, N.B.~Skachkov$^2$ \\
Joint Institute for Nuclear Research, Dubna, Russia\\
E-mail: (1) dmv@cv.jinr.ru, (2) skachkov@cv.jinr.ru
}

\maketitle

\begin{abstract}
\noindent
The number of \gpj and \zpj events suitable for a determination of gluon distribution function
 in a proton $f^g_p(x,Q^2)$ at Tevatron Run~II and during the low luminosity runs at future LHC experiments
are estimated
for various intervals of $x$ and $Q^2$. These numbers allow to extract $f^g_p(x,Q^2)$
in  new kinematic regions of 
$2\cdot 10^{-3}\leq x \leq 1.0$ with $1.6\cdot10^3\leq Q^2\leq2\cdot10^4 ~(GeV/c)^2$ at Tevatron and
of $2\cdot 10^{-4}\leq x \leq 1.0$ with $1.6\cdot10^3\leq Q^2\leq2\cdot10^5 ~(GeV/c)^2$ at LHC.
The contributions of background events in different $Q^2$ intervals are also given. 
\end{abstract}

\section{Introduction.}
\label{intro}

To study the possibilities of getting out the information about the gluon distribution
from inclusive photons \cite{NLO}
and from \gzpj events [2--5]
measured in $pp$ or $p\bar{p}$ collisions
we estimate here the numbers of events that can be collected
during the nearest 2--3 years of running with the planned luminosity at Tevatron Run~II and 
LHC (at $L=10^{33} s^{-1}cm^{-1}$) experiments as well as the contribution from the background events.
The process $pp(p\bar{p})\rightarrow \gamma/Z^0\, +\, 1\,jet\, + \,X$, we are studying,
is caused at the leading order by the following subprocesses:\\[-19pt]

%
%
~\\[-19pt]
\begin{eqnarray}
\hspace*{2mm} qg\to q+\gamma/Z^0 ~~(1a),  ~~~~~~~~ q\overline{q}\to g+\gamma/Z^0 ~~(1b).
\nonumber
\end{eqnarray}

The results presented here  are heavily based on the studies done in 
[2--5]
where the event selection criteria 
of  events with a clean \gzpj topology and most suitable for 
jet energy scale setting at Tevatron and LHC 
were developed. 
Our estimations are done mainly within the framework of
PYTHIA event generator \cite{PYT} and complemented by  detector response simulations
while studying a possibility of the background events rejection \cite{BKS_GLU,BS_MES_QG}
\footnote{The geometry of D0 and CMS detectors are used for all presented estimations.}.
%

%
\section{The background estimation.}
%

To estimate a background to the signal events, we have done 
by three generations (each of about 40 million events)
with a mixture of all QCD and SM subprocesses existing in PYTHIA 
including subprocesses (1a) and (1b).
%
%
Each event generation was done for three different values of
minimal transverse momentum of a hard subprocess 
%
%
$\pth$: 40, 70, 100 $GeV/c$ for Tevatron and 40, 100, 200 $GeV/c$ for LHC.

The background to the \gpj events 
is mainly caused by
the events with high $\Pt$ photons produced in the neutral decay channels 
of $\pi^0, \eta, \omega$ and $K^0_s$ mesons (``$\gamma\!-\!mes$'' events) 
%
and by
the events with the photons radiated from quarks (i.e. bremsstrahlung 
photons) in the next-to-leading order QCD subprocesses 
(``$\gamma\!-\!brem$'' events). 

The background may be also caused by ``$e^{\pm}$ events'' containing one jet and
$e^{\pm}$ as a direct photon candidate. The fraction of these events in the total background
turns out to be negligibly small after application of all selection cuts 
(including the cut on the missing transverse momentum in an event)
and the account of the tracker information.

The selection criteria of the \gpj events are described in detail in \cite{D0_Note,CMS_gpj,PartIV}.
In is worth mentioning  among them only two new ones, not used earlier in previous experiments. 
They leave only the events having small values of cluster (mini-jet)
transverse momenta, i.e. $\Pt^{clust}$, and the modulus of a vector sum of  transverse momenta 
of all detectable particles that are out of \gpj system, i.e. $\Pt^{out}$, 
by limiting them from above by $10~GeV/c$
\footnote{The selection criteria 
(mostly the two of them that limit $\Pt^{clust}$ and $\Pt^{out}$ \cite{D0_Note,CMS_gpj,CMS_zpj}
allow the events with a good \ptgj balance  to be selected because they 
provide an essential initial and final state radiation suppression,
i.e. a suppression of the contribution of the next-to-leading order diagrams.}.

The obtained after application of the selection cuts 
the signal-to-background ($S/B$) ratios are presented in Table \ref{tab:sb} 
\footnote{They are given without account of the contribution from 
the ``$e^\pm$ events'' \cite{D0_Note,CMS_gpj,PartIV}.}.
~\\[-7mm]
\begin{table}[h]
\small
\begin{center}
\caption{$S/B$ at Tevatron and LHC.}
\vskip 1mm
\begin{tabular}{|c||c|c|c||c|c|c|} \hline
\label{tab:sb}
$\pth~(GeV/c)$ & 40  & 70  & 100  & 40  & 100  & 200 \\\hline
$S/B$          & 2.9 & 8.8 & 21.1 & 4.1 & 22.4 & 47.0 \\\hline
\end{tabular}
\end{center}
\vskip-2mm
\end{table}

The estimations in PYTHIA have also shown
that most of ``$\gamma-brem$'' and ``$\gamma-mes$'' events ($80\%$ at least) originate from  
$qg\to qg$ and $qq\to qq$ scatterings with dominant contribution from the first subprocess 
($60-70\%$) \cite{PartIV}.

%
\section{Event rate estimation for gluon distribution determination at the Tevatron Run~II and LHC.}
%

\begin{figure}[htbp]
\vskip24mm
\hspace{10mm}
\includegraphics{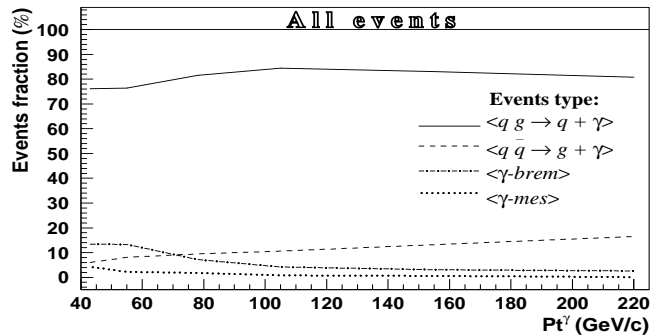}
\vskip 14mm
\caption[*]{The contributions of various event types to the total number of events as a function 
of $\Ptg$.}
\label{fig:all_ev}
\vskip-3.0mm
\end{figure}

The rates of the selected $q g\rrr \gamma q$ events 
are presented in Table \ref{tab:qg_tev} for the case of Tevatron (at integrated luminosity
$L_{int}=3 ~fb^{-1}$)  for different intervals of $Q^2=(\Pt^{\gamma})^2$ and
parton momentum fractions $x$. 
The corresponding rates of $q g\to \gamma q$ 
events for the case of LHC (calculated with $L_{int}=10 ~fb^{-1}$) are shown
in Table \ref{tab:B29_0}.
%
%
The fractions of each event type in
a given interval of $\Pt^{\tg}$ are presented in Fig.~\ref{fig:all_ev} 
($100\%$ are taken for all events) for a case of LHC. 

 \begin{table}[h]
\small
\begin{center}
\vskip-4mm
\caption{Numbers of ``$\com$'' events (divided by $10^3$)
in $Q^2$ and $x$ intervals for $L_{int}=3~fb^{-1}$. Tevatron.}
\label{tab:qg_tev}
\vskip0.1cm
\small
\begin{tabular}{|lc|r|r|r|r|r|r|}                  \hline
&$Q^2$          &\multicolumn{5}{c|}{ \hspace{-0.9cm} $x$ values of a parton} &All $x$  \\\cline{2-8}
&$(GeV/c)^2$    &.001--.005 &.005--.01 &.01--.05 &.05--.1 & .1--1.&.001--1.     \\\hline
&1600-2500\hmm  &  8.6      & 56.3     &245.2    &115.9   &206.6  &632.6  \\\hline
&2500-4900\hmm  &  0.4      & 13.5     &119.3    & 64.4   &123.0  &320.7 \\\hline
&4900-8100\hmm  &    0      & 0.2      & 17.9    & 13.5   & 27.4  & 59.0\\\hline
&8100-19600\hmm &    0      & 0.0      &  3.8    &  5.6   & 12.0  & 21.5\\\hline
\multicolumn{7}{c|}{}&{\bf 1 034}  \\\cline{8-8}
\end{tabular}
\vskip-3.7mm
\hspace*{69mm}({\bf $10^3\times $})
\vskip0mm
\vskip0.3cm
\caption{Numbers of ``$\com$'' events (divided by $10^3$)
in $Q^2$ and $x$ intervals at $L_{int}=10~fb^{-1}$. LHC.}
\label{tab:B29_0}
\small
\vskip0.1cm
\begin{tabular}{|lc|r|r|r|r|r|}  \hline
 & $Q^2$ &\multicolumn{4}{c|}{ \hspace{-0.9cm} $x$ values of a parton} &All $x$   
\\\cline{3-7}
 & $(GeV/c)^2$ & $10^{-4}$--$10^{-3}$ & $10^{-3}$--$10^{-2}$ &$10^{-2}$--
$10^{-1}$ & $10^{-1}$--$10^{0}$ & $10^{-4}$--$10^{0}$     \\\hline
&\hmm\hmm 1600-2500\hmm  &  830.7  & 2503.8  & 2577.6  &  221.3  & 6133.3   \\\hline
&\hmm\hmm 2500-5000\hmm  &  358.0  & 1497.9  & 1615.2  &  233.8  & 3704.8    \\\hline
&\hmm\hmm 5000-10000\hmm &   36.0  &  380.3  &  415.7  &  116.5  &  948.4     \\\hline
&\hmm\hmm 10000-20000\hmm&    1.9  &   84.8  &   98.9  &   46.1  &  231.7    \\\hline
&\hmm\hmm 20000-40000\hmm&    0.0  &   16.9  &   24.9  &   13.3  &   55.1  \\\hline
&\hmm\hmm 40000-80000\hmm&    0.0  &    2.9  &    5.4  &    3.8  &   12.0  \\\hline
\multicolumn{6}{c|}{}&{\bf 11 084}\\\cline{7-7}
\end{tabular}
\vskip-3.8mm
\hspace*{71mm}({\bf $10^3\times $})
\vskip-2mm
\end{center}
\end{table}

The estimations of event rates with a charm quark $g c\to \gamma c$  at Tevatron and LHC
are given in \cite{D0_Note,CMS_gpj}
\footnote{They give approximately 14 times less number of events.}.

The discrimination efficiencies of 
a single photon from the $\pi^0, \eta, K^0_s$ mesons (decayed via neutral channels) 
as well as those between quark and gluon jets 
allows to increase noticeably the purity of ``$\gamma^{dir}+jet$''
events \cite{PartIV,BS_MES_QG}.

Another possibility to extract a gluon distribution in a proton is a usage
of \zpj events with the subsequent $Z^0$ decay via leptonic ($l^+l^-$) channels. The 
selection criteria  guarantee practically complete suppression of the background events 
\cite{CMS_zpj}. 
The distribution of the number of the \zpj events, originated from subprocess  $qg\to Z^0+q$ 
(with the decay $Z^0\to\mu^+\mu^-,e^+e^-$) over $Q^2$ and $x$ intervals
are given in Table \ref{tab:B30} for the case of LHC at $L_{int}=20~fb^{-1}$. 

Fig.~2 shows a kinematic plot with the area that can be covered by studying 
the process $q g\rrr \gamma q$ at Tevatron and LHC.
The distribution of events  inside this area is given in Tables \ref{tab:qg_tev} and \ref{tab:B29_0}.
It is seen that at Tevatron (full line) 
and   
 at LHC (dashed line) it would be possible to study the gluon distribution on a good
statistics of \gpj events
in the region of small $x$ at values of $Q^2$ that are about by $1-2$ orders of magnitude higher than
those that are reached at HERA now.
It is also worth emphasizing that the $x-Q^2$ region covered by Tevatron has a common region 
with HERA at $0.05\lt x \lt 0.5$.
\begin{figure}[h]
\vskip64mm            
\hspace*{6mm} 
\includegraphics{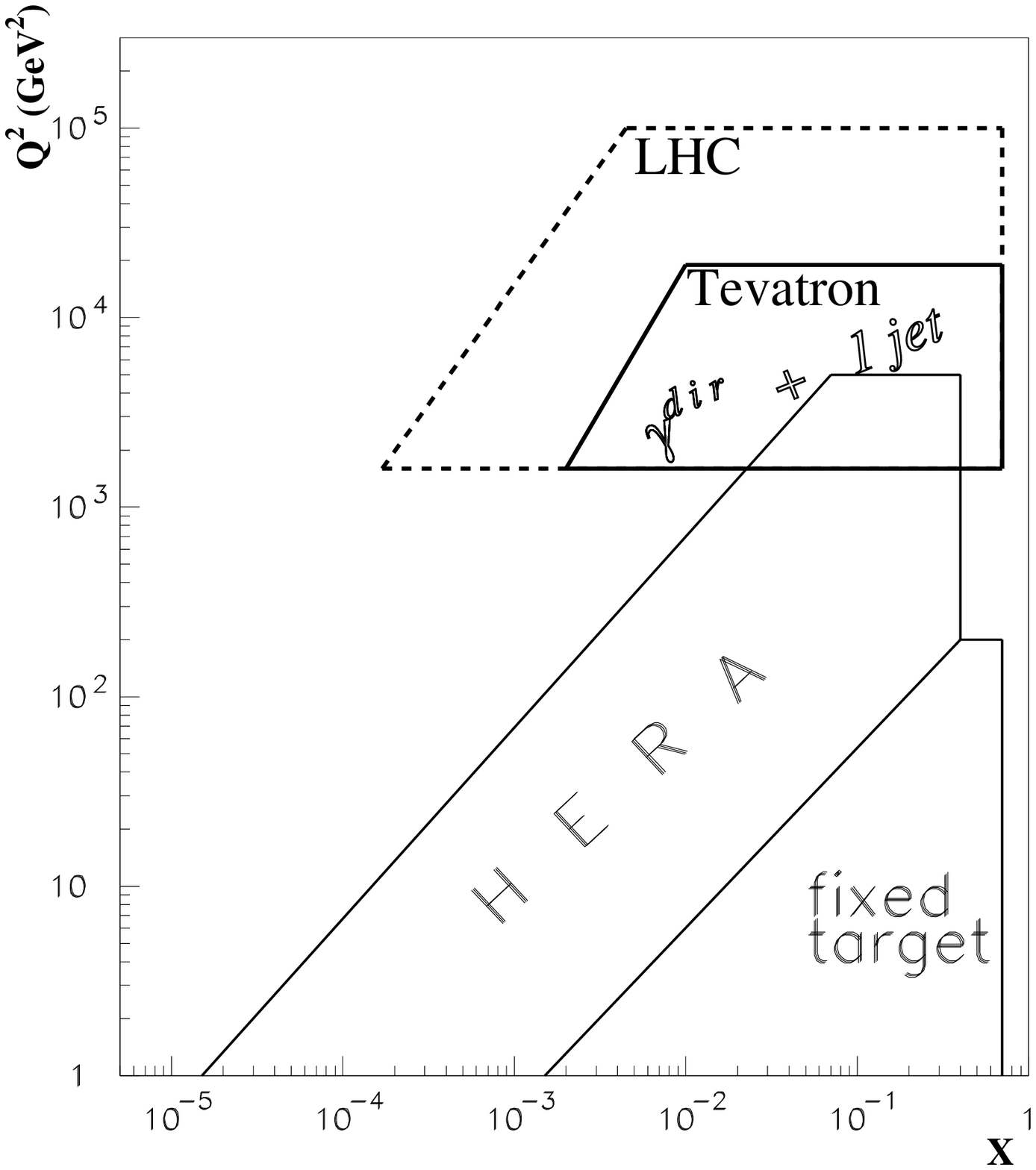}
\vskip-78mm
\label{fig:kinem}
\end{figure}
\vspace*{6mm}
\hspace*{56.3mm}
\parbox[r]{.52\linewidth}
{  Such a overlapping allows to carry out the cross-check of the
 $f^g_p(x,Q^2)$ measurements at Tevatron and HERA. 
The data from Tevatron, in their turn, will be important
for analogous cross-check of the $f^g_p(x,Q^2)$ measurements at LHC (see Fig.~2).
%
So, the data from Tevatron combined with ones from HERA and with the future data from LHC would allow to fulfill the QCD
analysis in $Q^2$ region varying in the wide interval of $10^2\lt Q^2 \lt 10^5~GeV^2$.
}

\vskip 8mm
{
\noindent 
{Figure 2: \normalsize {Kinematic region for $pp(\bar{p}) \to \gamma+jet+X$ process at 
Tevatron and LHC.}
}

\begin{table}[h]
\small
\begin{center}
\vskip-0.3cm
\caption{Numbers of ``$g q\to Z^0 + q$'' events (with $Z^0\to\mu^+\mu^-,e^+e^-$)  in 
$Q^2$ and $x$ intervals for $L_{int}=20 ~fb^{-1}$. LHC.}
\label{tab:B30}
\vskip0.1cm
\begin{tabular}{|lc|r|r|r|r|r|}                  \hline
 & $Q^2$ &\multicolumn{4}{c|}{ \hspace{-0.9cm} $x$ values of a parton} &All $x$ 
  \\\cline{3-7}
 & $(GeV/c)^2$ & $10^{-4}$--$10^{-3}$ & $10^{-3}$--$10^{-2}$ &$10^{-2}$--
$10^{-1}$ & $10^{-1}$--$10^{0}$ & $10^{-4}$--$10^{0}$     \\\hline
&\hmm\hmm  900-1600  \hmm  & 19563  & 55852  & 59257  &  2838  &137510 \\\hline 
&\hmm\hmm 1600-2500  \hmm  &  8627  & 38407  & 40262  &  2422  & 89718 \\\hline
&\hmm\hmm 2500-5000  \hmm  &  5108  & 37197  & 45256  &  4276  & 91837 \\\hline
&\hmm\hmm 5000-10000 \hmm  &   530  & 19866  & 25391  &  3368  & 49154 \\\hline
&\hmm\hmm 10000-20000\hmm  &    38  &  6735  & 11882  &  1968  & 20623 \\\hline
&\hmm\hmm 20000-40000\hmm  &    0  &  1324  &  3746  &  1097  &  6168  \\\hline
\multicolumn{6}{c|}{}&{\bf 395~010}\\\cline{7-7}
\end{tabular}
\end{center}
\vskip -1mm
\end{table}

\newpage
\noindent
{\bf Acknowledgments} \\[7pt]
We are greatly thankful to D.~Denegri, P.~Aurenche,
M.~Dittmar, M.~Fontannaz, J.Ph.~Guillet, M.L.~Mangano, E.~Pilon,
H.~Rohringer, S.~Tapprogge, H.~Weerts and J.~Womersley.

\end{document}